\documentclass[
    ,final            
  ]
  {aipproc}

\layoutstyle{6x9}

\renewcommand\XFMtitleblock{%
\XFMtitle
\let\XFMoldpar\par
\def\par{\XFMoldpar\def\par{\space
    for the VERITAS Collaboration\footnote{\protect\url{http://veritas.sao.arizona.edu/}}\XFMoldpar}}%
 \XFMauthors
 \let\par\XFMoldpar
 \XFMaddresses
 \XFMabstract
 \vspace{5pt}%
 \XFMkeywords
 \XFMclassification
}

\begin{document}

\title{VERITAS Limits on Dark Matter Annihilation from Dwarf Galaxies}


\author{J. Grube}{
  address={Astronomy Department, Adler Planetarium, Chicago, IL 60605, USA}
  ,altaddress={Enrico Fermi Institute, University of Chicago, Chicago, IL 60637, USA}
}
\begin{abstract}
Current cosmological models and data suggest the existence of a Cold Dark Matter (DM) component, however the nature of DM particles remains unknown. A favored candidate for DM is a Weakly Interacting Massive Particle (WIMP) in the mass range from 50 GeV to greater than 10 TeV. Nearby dwarf spheroidal galaxies (dSph) are expected to contain a high density of Dark Matter with a low gamma-ray background, and are thus promising targets for the detection of secondary gamma rays at very high energies (VHE, E $>$ 0.1 TeV) through the annihilation of WIMPS into SM particles. Presented here are recent VERITAS observations of dSph, including a deep exposure on Segue 1. Limits are derived for various annihilating and decaying dark matter particle models.
\end{abstract}

\maketitle


\section{Introduction}
The indirect search for very high energy (VHE) $\gamma$-rays resulting from the annihilation of Dark Matter (DM) particles into SM particles provides an important complement to that of direct searches for DM interactions and accelerator production experiments. Among theoretical candidates for the DM particle \cite{Bertone05}, a weakly interacting massive particle (WIMP) is well-motivated since it naturally provides the measured present day cold DM density \cite{Komatsu11}. Candidates for WIMP dark matter are present in many extensions of the SM of particle physics, such as supersymmetry (SUSY) \cite{Jungman96} or theories with extra dimensions \cite{Servant03}. In such models, the WIMPs either decay or annihilate into standard model particles, producing either a continuum of $\gamma$-rays with energies up to the DM particle mass, or monoenergetic $\gamma$-ray lines. 

Promising targets for indirect DM searches are nearby regions with high DM density. The Galactic Center is likely the brightest source of $\gamma$-rays resulting from DM annihilations, however the detected VHE $\gamma$-ray emission is coincident with the supermassive black hole Sgr A$^{*}$ and a nearby pulsar wind nebula \cite{Aharonian06}, motivating searches for DM annihilation in the Galactic Center halo where the VHE $\gamma$-ray background is expected to be significantly lower \cite{Abramowski11}. Alternatively, dwarf spheroidal galaxies (dSphs) of the Milky Way are attractive targets for DM searches. This is due to their relatively well-constrained DM profiles derived from stellar kinematics, proximity of 20--100 kpc, and general lack of active or recent star formation, suggesting a relatively low background from conventional astrophysical VHE processes \cite{Mateo98}. Here we present results for a search of DM annihilation signal from five dSphs observed with VERITAS. 

\section{Observations}
VERITAS is an array of four imaging atmospheric Cherenkov telescopes (IACTs), each 12 m in diameter, located at the Fred Lawrence Whipple Observatory in southern Arizona, USA. Each VERITAS camera contains 499 pixels (0.15$^{\circ}$ diameter) and has a field of view of 3.5$^{\circ}$. In the summer of 2009 the first telescope was moved to its current location in the array to provide a more uniform distance between telescopes, improving the sensitivity of the system \cite{Perkins09}. VERITAS is sensitive over an energy range of 100 GeV to 30 TeV with an energy resolution of 15\%--20\% and an angular resolution (68\% containment) of less than 0.1$^{\circ}$ per event. A source with a 1\% Crab Nebula flux can be detected by VERITAS in approximately 25 hours.

Table \ref{TableObs} lists the VERITAS observations of dSphs. Data-quality selection requires clear atmospheric conditions, based on infrared sky temperature measurements, and nominal hardware operation. The total exposures after dead time correction for each dSph are listed in table \ref{TableObs}. The observations of Draco, Ursa Minor, Willman 1, and Bo\"{o}tes 1 were taken before the first telescope was moved, while the Segue 1 data was taken in the current array layout. All data were taken during moon-less periods in wobble mode with pointings of 0.5$^{\circ}$ offset \cite{Aharonian01}. Data reduction followed the standard methods, yelding consistent results with two analysis packages \cite{Acciari08}.
\begin{table}
\begin{tabular}{lclccccc}
\hline
    \tablehead{1}{l}{b}{dSph\\}
  & \tablehead{1}{c}{b}{Dist.\\ (kpc)}
  & \tablehead{1}{c}{b}{Obs. Period\\}
  & \tablehead{1}{c}{b}{Exp.\\(hr)}
  & \tablehead{1}{c}{b}{Sig.\\}
  & \tablehead{1}{c}{b}{UL\\(cts)}
  & \tablehead{1}{c}{b}{E$_{\rm{Th}}$\\(GeV)}
  & \tablehead{1}{c}{b}{Flux UL\\(cm$^{-2}$ s$^{-1}$)}   \\
\hline
Draco      & 80 & 2007 Apr--May & 18.4 & -1.5 & 18.8  & 340 & 0.49 \\
Ursa Minor & 66 & 2007 Feb--May & 18.9 & -1.8 & 15.6  & 380 & 0.40 \\
Willman 1  & 38 & 2007--2008    & 13.7 & -0.1 & 36.7  & 320 & 1.17 \\
Bo\"{o}tes 1   & 62 & 2009 Apr--May & 14.3 &  1.4 & 72.0  & 300 & 2.19 \\
Segue 1	   & 23 & 2010--2011    & 47.8 &  1.4 & 102.5 & 300 & 0.80 \\
\hline
\end{tabular}
\caption{VERITAS observations of dwarf spheroidal galaxies (dSph).}
\label{TableObs}
\end{table}

\section{Flux Upper Limits}
\begin{figure}
\includegraphics[height=.25\textheight]{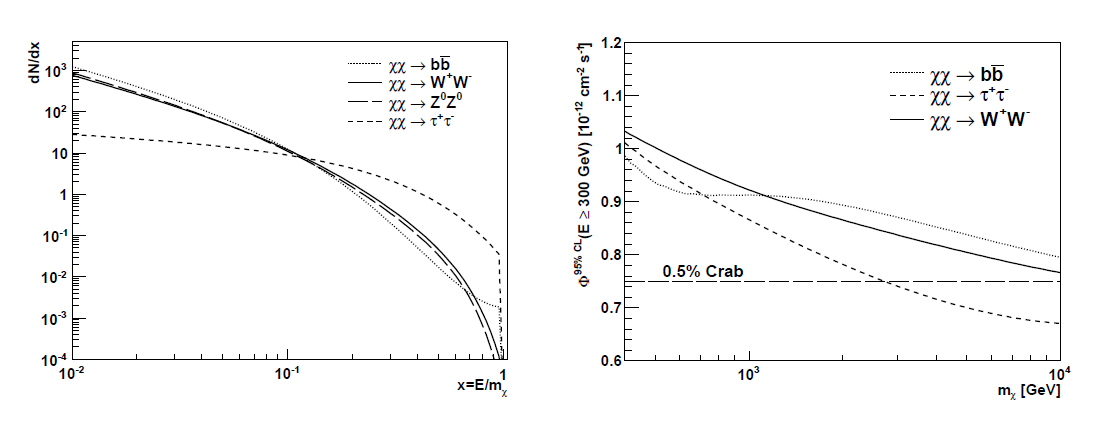}
\caption{{\bf Left:} Dark Matter annihilation spectra for four different final state products ($W^{+}W^{-}$, $Z^{0}Z^{0}$, $b\bar{b}$, and $\tau^{+}\tau^{-}$), extracted from PYTHIA 8.1. {\bf Right:} Upper limits (95\% CL) on the integrated $\gamma$-ray flux above $E_{min} = 300$ GeV from the VERITAS observations of Segue 1 considering dark matter particle annihilation/decay for three different channels: $W^{+}W^{-}$, $b\bar{b}$, and $\tau^{+}\tau^{-}$. \cite{Aliu12}.}
\label{Fig1}
\end{figure}

As listed in table \ref{TableObs}, no significant VHE $\gamma$-ray signal is detected from the dSph observed by VERITAS. Upper Limits (UL) on the integrated flux are calculated based on the assumption that the dominant source of $\gamma$-rays is DM annihilation. The $\gamma$-ray differential energy spectrum from DM particle annihilation depends on the DM model, in particular on the branching ratios to the final state particles. In almost every channel (excepting the $e^{+}e^{-}$ and $\mu^{+}\mu^{-}$ channels), the $\gamma$-ray emission mostly originates from the hadronization of the final state particles, with the subsequent production and decay of neutral pions. Annihilation to three different final state products is considered independently of the dark matter model, in each case with a 100\% branching ratio. Figure \ref{Fig1} (left) shows the annihilation spectra for the channels $W^{+}W^{-}$, $b\bar{b}$, and $\tau^{+}\tau^{-}$. For each channel, the $\gamma$-ray spectrum has been simulated with the particle physics event generator PYTHIA 8.1. Figure \ref{Fig1} (right) shows the ULs (95\% CL) on the integrated flux above $E_{min} = 300$ GeV from VERITAS observations of Segue 1 as a function of the DM particle mass. The integrated flux ULs range between 0.3\% and 0.7\% of the Crab Nebula integral flux, depending on the DM particle mass.

\section{Upper Limits on the DM Annihilation Cross-Section}
The $\gamma$-ray flux from the annihilation of DM particles in a spherical DM halo is determined in part by J($\Delta\Omega$), the squared DM density integrated along the line of sight and over a solid angle given by the size of the signal search region defined in the VERITAS analysis. The estimate of J($\Delta\Omega$) requires a modeling of the dSph DM profile, which for the observations of Segue 1 an Einasto profile \cite{Navarro10} is applied, resulting in J($\Delta\Omega$) $= 7.7 \times 10^{18}$ GeV$^{2}$ cm$^{-5}$ sr. The upper limit (UL) on the total velocity-weighted annihilation cross-section $\langle \sigma \nu \rangle^{95\%  \rm{CL}}$ is then calculated using the UL on $\gamma$-like events N$_{\gamma}^{95\% \rm{CL}}$, the total observation time T$_{\rm{obs}}$ and the effective area A$_{\rm{eff}}$ for the dataset, and the DM mass m$_{\chi}$:
\begin{equation}
\langle \sigma \nu \rangle^{95\%  \rm{CL}} = \frac{8\pi}{\bar{J}(\Delta\Omega)} \times \frac{N_{\gamma}^{95\% \rm{CL}} m_{\chi}^{2}}{T_{\rm{obs}} \int_{0}^{m_{\chi}} A_{\rm{eff}} \frac{dN_{\gamma}}{dE}dE}
\end{equation}

The main uncertainties in the DM limits computed here are from the modeling of the DM density profile. Unlike the classical dSphs, there is no high-statistics star sample for Segue 1. This prevents an accurate modeling of the DM distribution. Assuming an Einasto profile \cite{Navarro10}, the systematic uncertainties in the dark matter profile modeling can change the astrophysical factor, and hence they scale up or down the limits by a factor of 4 at the 1 $\sigma$ level \cite{Essig10}. Future spectroscopic surveys should increase the Segue 1 star sample and eventually reduce the uncertainties on its dark matter content.

\begin{figure}
\includegraphics[height=.25\textheight]{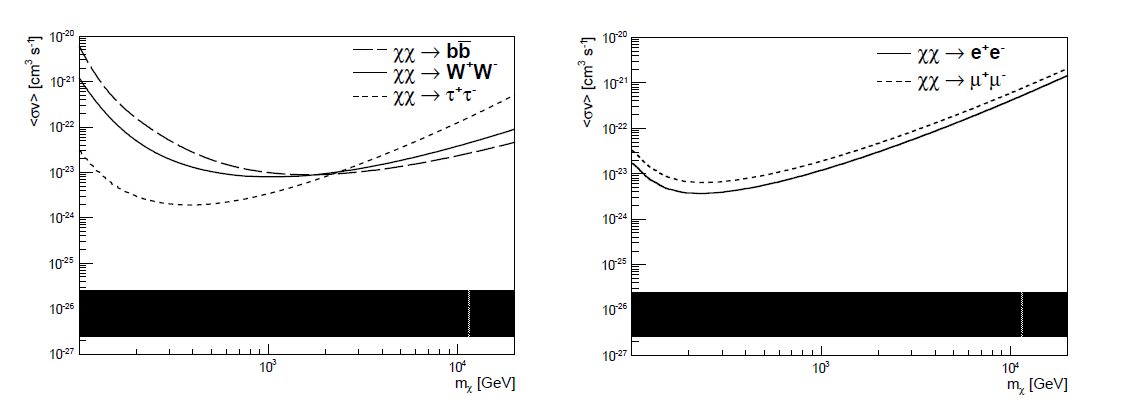}
\caption{
Upper limits (95\% CL) from VERITAS observations of Segue 1 on the WIMP velocity-weighted annihilation cross-section $\langle \sigma \nu \rangle$ as a function of the WIMP mass, considering different final state particles. The grey band area represents a range of generic values for the annihilation cross-section in the case of thermally produced dark matter. {\bf Left:} hadronic channels $W^{+}W^{-}$, $b\bar{b}$, and $\tau^{+}\tau^{-}$. {\bf Right:} leptonic channels $e^{+}e^{-}$ and $\mu^{+}\mu^{-}$ \cite{Aliu12}.}
\label{Fig2}
\end{figure}

\begin{theacknowledgments}
This research is supported by grants from the U.S. Department of Energy Office of Science, the U.S. National Science Foundation and the Smithsonian Institution, by NSERC in Canada, by Science Foundation Ireland (SFI 10/RFP/AST2748) and by STFC in the U.K. We acknowledge the excellent work of the technical support staff at the Fred Lawrence Whipple Observatory and at the collaborating institutions in the construction and operation of the instrument.
\end{theacknowledgments}


\end{document}